\documentclass{aa}  

\usepackage{graphicx}
\usepackage{txfonts}
\begin{document}

   \title{Long-term radial-velocity variations of the Sun as a star: \\ 
            The HARPS view}
              \titlerunning{Long-term RV variations of the Sun as a star}
  \authorrunning{A. F. Lanza et al.}


   \author{A.~F.~Lanza
          \inst{1}
          \and
          P. Molaro\inst{2}
          \and L.~Monaco\inst{3}
          \and R.~D.~Haywood\inst{4,5}
          }

   \institute{INAF-Osservatorio Astrofisico di Catania, Via S.~Sofia, 78 - 95123 Catania, Italy\\
              \email{nuccio.lanza@oact.inaf.it}
         \and
             INAF-Osservatorio Astronomico di Trieste, Via G.~B.~Tiepolo, 11 - 34143 Trieste, Italy\\
             \email{molaro@oats.inaf.it}
             \and
             Universidad Andres Bello, 
Departamento de Ciencias Fisicas,
Republica 220, Santiago, Chile \\ 
             \email{lorenzo.monaco@unab.cl}
             \and
             School of Physics and Astronomy, University of St. Andrews, St. Andrews KY16 9SS, UK\\
             \email{rdh4@st-andrews.ac.uk}           
             \and 
              Harvard-Smithsonian Center for Astrophysics, 60 Garden Street, Cambridge, MA 02138, USA\\
             \email{raphaelle.haywood@cfa.harvard.edu}             
             }

   \date{Received ... ; accepted ...}

 
  \abstract
   {Stellar radial velocities play a fundamental role in the discovery of extrasolar planets and the measurement of their physical parameters as well as in the study of stellar physical properties.}
   {We investigate the impact of the solar activity on the radial velocity of the Sun using the HARPS spectrograph to obtain measurements that can be directly compared with those acquired in the extrasolar planet search programs.}
   {We use the Moon, the Galilean satellites, and several asteroids as reflectors to measure the radial velocity of the Sun as a star  and correlate it with disc-integrated chromospheric and magnetic indexes of solar activity that are similar to stellar activity indexes. We discuss in detail the systematic effects that affect our measurements and the methods to account for them.}
   { We find that the radial velocity of the Sun as a star is positively correlated with the level of its chromospheric activity at $\sim 95$ percent significance level. The amplitude of the long-term variation measured in the 2006-2014 period is $4.98 \pm 1.44$~m/s, in good agreement with model predictions. The standard deviation of the residuals obtained by subtracting a linear best fit is 2.82~m/s and is due to the rotation of the reflecting bodies and the intrinsic variability of the Sun on timescales shorter than the activity cycle. A correlation with a lower significance  is detected between the radial velocity and the mean absolute value of the line-of-sight photospheric magnetic field flux density. }
   {Our results confirm similar correlations found in other late-type main-sequence stars and provide support to the  predictions of radial velocity variations induced by stellar activity based on current models.}


   \maketitle
%

\section{Introduction}
Extrasolar planets were first detected by measuring the radial velocity (hereafter RV) oscillations of the host stars induced by their orbital motion  \citep[cf.][]{MayorQueloz95}. This method  allows to measure the orbit eccentricity and the true mass of a planet when it transits across the disc of the host star thus allowing us to measure the inclination of its orbital plane. The new frontier is the detection of Earth-mass planets and the measurement of their mass \citep[cf. ][]{Dumusqueetal14,Haywoodetal14} which is made difficult by the apparent RV variations of the host star produced by its p-mode oscillations, surface convection, and magnetic activity.  While the first two effects can be mitigated by averaging several observations on the same night \citep{Dumusqueetal11} { or on consecutive nights  \citep{Meunieretal15}}, magnetic activity is much more complex to take into account and can mimic a planetary signal on a variety of timescales \citep[e.g.,][]{Santosetal14}. In particular, the detection of planets with orbital periods of several years or decades can be hampered by the RV modulations associated with stellar activity cycles \citep[e.g.,][]{Santosetal10,Lovisetal11,GomesdaSilvaetal12,Caroloetal14,Moutouetal15}.

A detailed understanding of this phenomenon and of the correlations between the RV and the magnetic activity indexes can be achieved by observations of the Sun as a star because we have simultaneous information on the disc position and parameters of the individual active regions. This  allows us to test models that predict activity-induced RV perturbations from  resolved images of the disc of the Sun such as those introduced by \citet{Lagrangeetal10} or \citet{Meunieretal10a}. They consider the effect of surface brightness inhomogeneities, either sunspots or faculae, that produce a distortion of the spectral line profiles that depend on their contrast, filling factor, position, and the rotation velocity of the Sun. In addition to this effect, there is a  quenching of convective motions by the magnetic fields that affects the line convective blueshifts. Which effect is prevailing depends on several factors, notably the  filling factor of  the active regions and the projected rotational velocity of the star. Moreover, while the effect of the brightness inhomogeneities can be either positive or negative depending on the sign of their contrast and the location on the receding or the approaching half of the solar disc, the magnetic quenching of convection always produces an apparent redshift \citep[e.g.,][]{Lanzaetal10,Lanzaetal11}. 

\citet{Meunieretal10a} found that the main effect  on the  Sun-as-a-star RV measurements is produced by the quenching of the local convective flows in active regions.  This leads to an apparent redshift in the synthetic RVs that appears to be positively correlated with activity indexes, such as the chromospheric Ca~II~H\&K line core emission \citep{MeunierLagrange13}. 

To confirm these model predictions, we need RV measurements of the Sun as a star that extend over a significant part of a solar activity cycle in order to sample the variations at different activity levels. Long-term sequences spanning more than 35 years have been acquired to study the global oscillations of the Sun, but they are based on the measurement of the Doppler shift of a single or a few spectral lines. For example,  \citet{RocaCortesPalle14} found a clear anticorrelation between the RV variation as measured from the KI~769.9~nm~line and the international sunspot number. These observations are not comparable to the stellar ones which are based on the simultaneous measurements of thousands of photospheric spectral lines to increase precision as, e.g.,  in the case of the HARPS spectrograph \citep{Mayoretal03}. Other investigations support this conclusion by confirming that the long-term RV variations of the Sun as a star depend on the specific spectral range considered \citep{Jimenezetal86,McMillanetal93,DemingPlymate94}. 
{\citet{Meunieretal10b} extrapolated the RV variations as measured by MDI/SoHO in the Ni line at 676.8 nm to obtain the solar disc-integrated RV variations along  cycle 23. They confirmed the dominant role of the suppression of convective line-shifts in magnetized regions as predicted by \citet{Meunieretal10a}.}

Instruments to integrate  light over the solar disc and feed high-resolution spectrographs for star-like RV measurements are under construction or just beginning their commissioning phase \citep[Glenday et al. 2015, in prep.;][]{Dumusqueetal15,Strassmeieretal15}. Therefore, we adopt a different approach that exploits the optical solar spectrum as reflected by an asteroid, a Jupiter satellite, or a small area on the Moon \citep{Zwitteretal07,MolaroCenturion11,MolaroMonai12}. This avoids the effects of the differential extinction across the disc of the Sun in the Earth atmosphere that can reach up to tens of m/s when integrating the flux over the solar disc \citep[cf. ][]{McMillanetal93}.

\section{Observations}
 We searched the ESO HARPS archive looking for RV observations of asteroids, the Galilean satellites, or the Moon, and found a good deal of measurements acquired between 2006 and 2014 (cf. Table~\ref{table1})\footnote{{\tiny https://www.eso.org/sci/facilities/lasilla/instruments/harps/tools/archive.html}}. The Vesta time series acquired between 29 September and 7 December 2011 gives information on the impact of solar activity on the intra-night or daily variability of the RV of the Sun as a star and is investigated elsewhere \citep{Haywoodetal15}. Here we use only a few measurements of that dataset to complement our series because we are interested in variations over a longer timescale. 
 
Individual HARPS spectra  cover the  range 378-691 nm with a gap between 530 and 533 nm. The instrument has a fixed configuration with a spectral resolution of $\lambda /\Delta \lambda \sim 120000$. The exposure time ranges from 30 to 3600~s according to the apparent magnitude of the target, thus giving  a {photon noise error} from $\sim 0.2$ to $\sim 1.0$~m/s. HARPS is in vacuum and thermally isolated and it is equipped with an image scrambler which provides  a uniform spectrograph-pupil illumination thus allowing very precise and stable RV measurements. 
The intra-night and long-term stability {stays within 1.0 m/s for bright targets} and it is  checked by means of a reference Th-Ar lamp whose spectrum is occasionally sent on a second fiber for a simultaneous comparison with the target spectrum.  The RV values are obtained through the HARPS Data Reduction System (DRS) by cross-correlating the source spectrum with a template obtained from the high-resolution Fourier Trasform Spectrometer solar spectrum by \citet{Kuruczetal84}. Its zero point is uncertain by $\approx 100$~m/s that explains the offset of about 100 m/s that we have in our measurements \citep[see also][]{MolaroMonai12}. An accurate estimate is not possible because the Sun is always active, thus we decided to leave it in our sequence because it is a constant that does not affect in any way our investigation of the  solar RV variations. 

\begin{figure}
   \centerline{
   \includegraphics[width=10cm,angle=0]{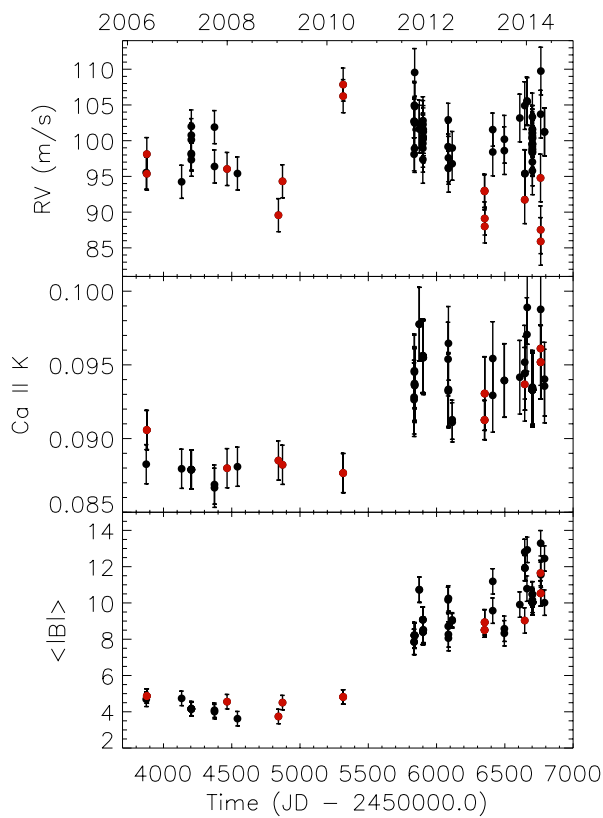}} 
\caption{Top panel: Solar RV as derived by observing different reflecting bodies vs. the time in JD on the lower axis and in years on the top axis, respectively. The errors are derived in Sect.~\ref{rv_uncertainty}. The red dots mark measurements that are affected by the rotation of the reflecting body or by other systematics and that are excluded from further analysis. Middle panel: The chromospheric Ca~II~K index corresponding to our RV measurements vs. the time (see Sect.~\ref{activity_index}). The red colour marks the excluded data points as in the top panel, while the errors are derived in Sect.~\ref{diff_persp}. Lower panel: the same as the middle panel, but for the total mean magnetic flux introduced in Sect.~\ref{activity_index} and whose errors are derived in Sect.~\ref{diff_persp}.    } 
              \label{timeseries}
    \end{figure}

The RV value given by the DRS is successively corrected for the Doppler shifts introduced by the orbital motion of the reflector with respect to the Sun and to the Earth  \citep[cf.][]{Zwitteretal07}, calculated by means of the NASA JPL Horizon ephemerides\footnote{Available through: http://ssd.jpl.nasa.gov/horizons.cgi}.  The relativistic Doppler shift is not considered by Horizon, although light aberration is included, given its direct effect on the apparent position of the bodies (Giorgini, priv. comm.). Therefore, we compute  the relativistic Doppler shift as explained in Appendix~\ref{appAA}, where we provide details on the procedure applied to compute star-like RV for the Sun. 
{Our RV time series is plotted in the top panel of Fig.~\ref{timeseries}, where we plot also the timeseries of the activity indexes as defined in Sect.~\ref{activity_index}. The red colour marks measurements affected by the rotation or the non-uniform albedo of the reflecting body (see Sect.~\ref{results}), while the measurement errors are derived in Sect.~\ref{rv_uncertainty} in the case of the RV and in Sect.~\ref{diff_persp} for the activity indexes, respectively.}

{In Table~\ref{table1} from left to right, we list the reflecting body, the epoch of each observation, its exposure time $\tau$, the measured RV of the Sun referred to its barycentre { (including the relativistic correction)}, the relativistic correction to the classic Doppler shift $\Delta V_{\rm rel}$ (cf. Appendix~\ref{appAA}), the time lag $t^{\prime}-t$ accounting for the different directions along which the Sun is seen from the reflecting body and the Earth (cf. Sect.~\ref{diff_persp}), the chromospheric Ca~II~K index, and the mean total magnetic field $<|B|>$ as defined in Sect.~\ref{activity_index}. The epoch of each observation in Table~\ref{table1}   is not the beginning of the exposure, but it is the time weighted according to the received stellar flux along the exposure as measured by the HARPS exposure meter. 

Most of the measurements obtained with the Moon are average values over sequences extended for several hours \citep[e.g.][]{Molaroetal13}. In Fig.~\ref{moon_sequence} we show an example of such lunar sequences. It was obtained on 5 July 2012 and its individual RV measurements    show a rather flat behavior with a mean of 98.3 m/s -- consider the typical offset of about 100 m/s -- and an  rms  of 0.83 m/s. The latter is mainly produced by the    5-min solar oscillations illustrating  the sub m/s precision of HARPS observations. On  4 and 6 January 2014, \citet{Molaroetal15} collected series of RV measurements by observing Ganymede and Europa, just before the beginning and after the end of the opposition surge effect that clearly affected the Europa data collected on  5 January. Therefore, we limit ourselves to only those parts of the dataset  not affected by this effect.  { Ganymede data on 4 January 2014 were obtained by HARPS-N, the twin spectrograph mounted at Telescopio Nazionale Galileo in La Palma.}
   \begin{figure}
   \centerline{
   \includegraphics[width=8.5cm,height=6.5cm,angle=0]{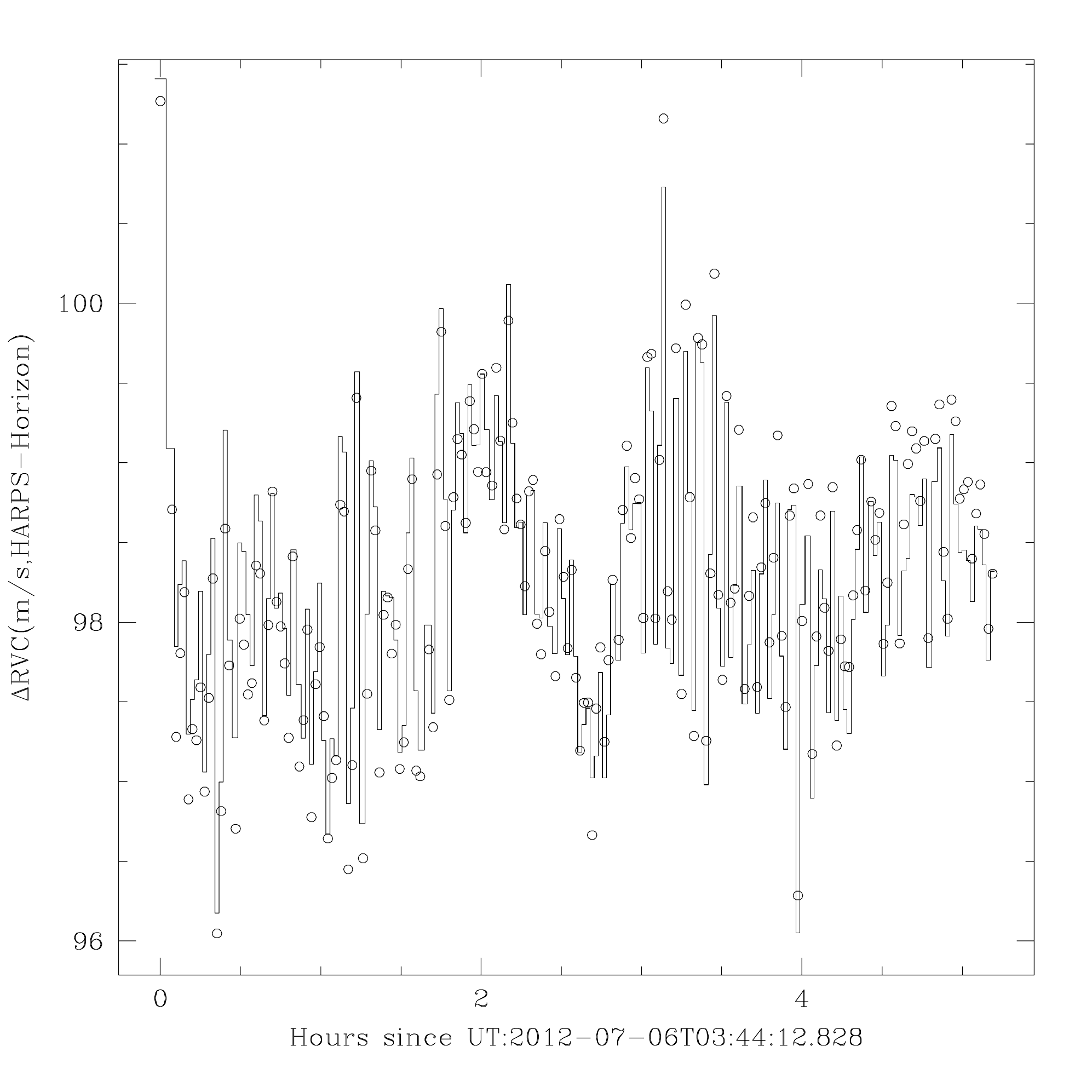}} 
   \caption{Solar RV  vs. the time for a sequence of 196  Moon measurements spanning  about 4.5 hours of total observation on 5 July 2012. The radial velocities are corrected by the radial motion components of the Moon relative to the observer and to the Sun as described in the text, but without rotational and relativistic corrections which are added in Table~\ref{table1}.  } 
              \label{moon_sequence}
    \end{figure}

Regarding the asteroids, in most of the cases, the exposure times are of 600~s or longer, thus averaging out the 5-min p-mode oscillations that were clearly detected in the Moon sequences \citep[cf.~Fig.~\ref{moon_sequence} and Fig.~4 in][]{Molaroetal13}. The standard deviations of the RV measurements obtained in the course of the same night are within $\sim 1.6-1.8$ m/s, except for those acquired with Vesta and Ceres that are  affected by the rotation of the asteroids (see Sect.~\ref{asteroid_rotation}). This is similar to the level of variability found on similar timescales in Sun-like stars \citep{Dumusqueetal11}. 

\subsection{Effects of the reflector rotation}
\label{asteroid_rotation}

The RV variation produced by the rotation of the Moon has been corrected using an adequate observing strategy. Specifically, most of the Moon spectra were acquired close to the centre of the Moon's disc, therefore, only the rotation velocity component in the direction of the Sun was relevant, while that in the direction of the Earth observer was negligible. 

In the case of the asteroids or the Galilean satellites, we receive a disc-integrated spectrum that is affected by the rotation of the body and the non-uniform illumination by the Sun. In the case of a body of spherical shape with a uniform albedo, the Doppler shift induced by its rotation averages to zero only if the inclinations of its rotation axis to the direction of the Sun and to the observer on the Earth are the same. If this is not the case, there is a net Doppler shift because  the opposite contributions coming from the receding and the approaching parts of the illuminated disc do not balance exactly with each other, as shown by \citet{LanzaMolaro15}. The amplitude of this effect can reach up to 2.6 m/s for Pallas whose spin axis is inclined by $\sim 16^{\circ}$ to the plane of the ecliptic, but it is smaller than 0.06 m/s for Ceres or Vesta having an inclination to the ecliptic greater than $60^{\circ}$ as well as for the Galileian satellites. {In addition to the inclination of the spin axis to the plane of the ecliptic, the RV variation  depends on the phase angle of the reflecting body and on the difference between its heliocentric ecliptic longitude and that of the Earth; therefore, it is modulated on timescales of weeks or months. }

Another effect occurs when the surface of the reflecting body has a non-uniform albedo as indicated in the case of Ceres and Vesta by the rotational modulation of their optical flux and the resolved images of their surfaces.  A dark spot of albedo produces a reduced continuum in the locally reflected spectrum of the Sun. The local spectra reflected by the different portions of the disc are Doppler shifted according to their projected  rotational velocities and the lower continuum in the dark spot  produces a bump in the spectral line profiles when we integrate the spectrum over the whole visible disc, akin to the case of a spotted star \citep[cf. Fig.~1 in][]{VogtPenrod83}. Assuming a continuum flux reduced by a factor $f_{\rm c} < 1$ with respect to that without the spot, the maximum RV perturbation is of the order of $2 (1-f_{\rm c}) V_{\rm eq}$, where the factor $2$  accounts for the worst case, i.e., when the rotational velocity components towards the Sun and the observer are equal to the equatorial rotation velocity of the asteroid $V_{\rm eq}$. It reaches up to $V_{\rm eq} \sim 90$ m/s in the case of Vesta or Ceres \citep[cf. Table~1 in][]{LanzaMolaro15}. Given that $0.90 \la f_{\rm c} \la 0.98$, RV perturbations of several m/s are possible as a consequence of albedo inhomogeneities on those fast-rotating asteroids having periods of $P_{\rm rot}=5.342$ and $9.075$ hr, respectively. 

A precise calculation of the effect is made difficult by our limited knowledge of the surface features of the different asteroids and by the variation of the angles between their spin axis and  the directions to the Sun and the Earth along their orbit. \citet{Haywoodetal15} makes an empirical investigation in the case of Vesta and find that a RV rotational modulation with a semi-amplitude of $\sim 2.39 \pm 0.55$ m/s has been induced by this effect  between 2011 September 11 and 2011 December 7 . In the case of the data points of Vesta in Table~\ref{table1}, we find a maximum standard deviation of $\sim 3.38$~m/s in the course of the same night, that is compatible with  their results and the very rough estimate given above.  { The amplitude is larger by a factor of $\approx 2$ in the case of Ceres \citep{Molaroetal16}}, while a smaller amplitude by a factor of $3-10$ is expected in the case of the other asteroids in our sample, owing to their smaller radii and longer rotation periods \citep[cf.][]{LanzaMolaro15}. For the Galilean satellites, the effect can be larger than $\sim 1$~m/s only for Io because it has the shortest rotation period. 


\subsection{Solar activity indexes}
\label{activity_index}

We use the normalized emission flux measured in a 0.1 nm bandpass centred in the core of the solar Ca II K line (central wavelength 393.3 nm) as a proxy for the solar chromospheric activity level \citep{KeilWorden84,Keiletal98,MeunierLagrange13}\footnote{Data are available from: http://nsosp.nso.edu/cak\_mon}. The typical relative accuracy of this index is $\sim 0.6$ percent. Furthermore, we consider the magnetic field averaged over the disc of the Sun as another proxy for its activity to allow a comparison with the results of \citet{DemingPlymate94} who used the mean absolute magnetic field over the solar disc as measured from Kitt Peak magnetograms. That time series terminated on 21 Sept 2003 and was then continued to date by the SOLIS Vector Spectromagnetograph (hereafter SOLIS~VSM). 

The mean field values are determined from daily measurements of the line-of-sight magnetic field flux density observed with 1 arcsecond 
pixels and averaged over the full disc. The solar line-of-sight magnetic field is measured with full-disc  
Fe I 630.15 nm (Land\'e factor of the line $g_{\rm eff}= 1.667$) longitudinal photospheric magnetograms \citep{Jonesetal02}\footnote{Data are available from: http://solis.nso.edu/vsm/vsm\_mnfield.html}. The mean net magnetic flux is the sum of all the measurements divided by the number of pixels on the disc. The mean total magnetic 
flux is the average of the absolute value of the pixel measurements. Only pixels within 0.99 solar radii from disc centre and with absolute values greater 
than 0.2~G are included in the computation. The mean net magnetic flux during the time interval of our RV observations spanned between $-0.89$ and $1.0$~G, while the range of the mean total flux was between 3.62 and 13.29~G. The mean standard deviations of the mean net and total fluxes are both of $\sim 0.017$~G.  

The variations of the chromospheric index and of the total magnetic flux vs. the time are plotted in Fig.~\ref{solar_cycle}. We see that the variations of the indexes are rather complex with the activity level of the Sun showing two relative maxima during cycle 24. Therefore, we cannot separate levels of low and high activity by simply considering the phase of the cycle as in the case of a single, well-defined maximum. 
\begin{figure}
   \centerline{
   \includegraphics[width=10cm,height=10cm,angle=0]{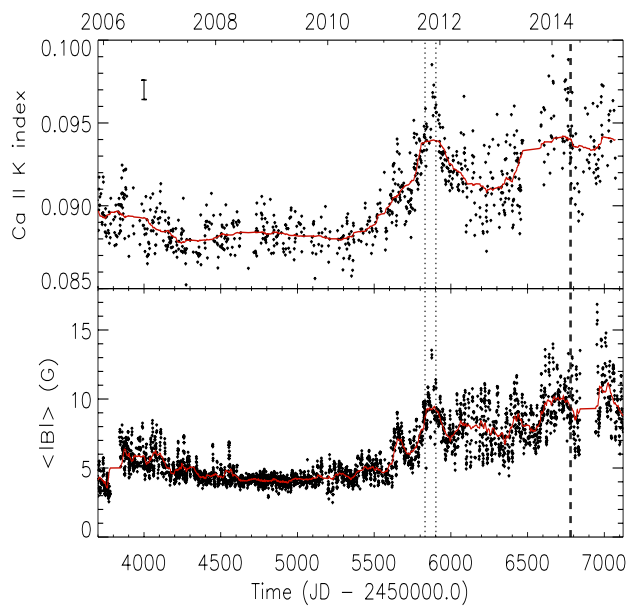}} 
\caption{Top panel:  The Ca II K chromospheric index (black dots) vs. the time in JD on the lower axis and in years on the top axis. The  continuous red line is obtained by smoothing the data with a time window of 60 days. The vertical dotted and dashed lines mark the time intervals during which measurements of the RV of the Sun as a star were acquired by \citet{Haywoodetal15} and \citet{Dumusqueetal15}, respectively. Lower panel: The same as the upper panel, but for the total magnetic flux as measured by SOLIS VSM. }
              \label{solar_cycle}%
\end{figure}
%


\subsection{Effect of the different lines of sight to the Sun}
\label{diff_persp}
In general,  the Sun is seen from different lines of sight from the Earth and the reflector,  therefore we must account for the time lag introduced by this effect when dealing with solar activity proxies. For an RV observation made at the time $t$, we compute the angle $\alpha$ between the directions  to the reflector and to the Earth as seen from the centre of the Sun. Assuming that the Ca~II~K emission comes from the faculae on the Sun and that they are rotating on average with the angular velocity $\Omega_{\rm AR}$ corresponding to a latitude of $\sim 20^{\circ}$, we compute the effective time $t^{\prime} = t - \alpha/ \Omega_{\rm AR}$ when the Sun is viewed by an Earth observer at the same rotation phase as viewed by an observer on the reflector at time $t$.  Therefore, we compute the Ca II K index to be associated to the RV measurement at time $t$ by linearly interpolating the  chromospheric index time series at the time $t^{\prime}$. The same method is used to compute the mean net and total magnetic fluxes  corresponding to a given RV measurement.The difference $t^{\prime}-t$ ranges from $-6.13$ to $7.10$ days with a mean value $\langle | t^{\prime} - t |\rangle \simeq 3.42$ days. It is much longer than the Sun-reflector-Earth light travel time that is always shorter than $\sim 2$ hours. Therefore, we neglect the light travel time when accounting for the different lines of sight. 

{ The time lag $t^{\prime}-t$ introduces an additional error because of the intrinsic variations of the chromospheric and magnetic indexes over that timescale. To quantify it, we compute their average standard deviation considering time intervals between 4 and 8 days for the period of high activity (Ca II K index greater than 0.092) and find values between 0.0020 and 0.0025 for the Ca II K index and  $0.45$ and $0.70$~G for the total magnetic flux, respectively. These  are significantly larger than the typical measurement errors and their mean values will be used as statistical errors of the activity indexes in our analysis.  }
 \begin{table*}
 \caption{HARPS radial velocity measurements of the Sun as a star. }
 \begin{tabular}{lrrrrrrrrrrrr}
 \hline
  & & & & & & & & & \\
  Object & year & month & day & hour & min & sec & $\tau\,\,$ & RV~~ & $\Delta V_{\rm rel}$ & $t^{\prime}-t$ & Ca~II~K & $<|B|>$  \\
   &  & & &  & & & (s) & (m/s) & (m/s) & (day) & & (Gauss) \\
   &  & & & & & & & & & & & \\
\hline 
&  & & & & & & & &  & & &  \\
\object{Europa} &   2006 &    5 &   18 &    3 &   56 &     0.50 &    180 &     95.54 &    -2.500 &     0.945 &    0.08826 &     4.687 \\ 
\object{Ceres} &   2006 &    5 &   22 &    9 &   36 &    12.23 &    300 &     95.37 &    -1.670 &    -4.940 &    0.09058 &     4.867 \\ 
Ceres &   2006 &    5 &   22 &    9 &   47 &     8.35 &    900 &     98.12 &    -1.671 &    -4.941 &    0.09058 &     4.867\\ 
\object{Vesta} &   2007 &    2 &    1 &    8 &   39 &    55.59 &    900 &     94.24 &    -2.756 &    -6.129 &    0.08795 &     4.738 \\ 
\object{Ganymede} &   2007 &    4 &   13 &    9 &   32 &    14.58 &     60 &     98.06 &    -1.378 &    -3.698 &    0.08789 &     4.158 \\ 
Ganymede &   2007 &    4 &   13 &    9 &   33 &    46.02 &     60 &     97.33 &    -1.369 &    -3.698 &    0.08789 &     4.158 \\ 
Ganymede &   2007 &    4 &   13 &    9 &   35 &    17.04 &     60 &    100.13 &    -1.369 &    -3.698 &    0.08789 &     4.158 \\ 
Ganymede &   2007 &    4 &   13 &    9 &   36 &    48.00 &     60 &    100.77 &    -1.369 &    -3.698 &    0.08789 &     4.158 \\ 
Ganymede &   2007 &    4 &   13 &    9 &   38 &    18.99 &     60 &    101.99 &    -1.369 &    -3.698 &    0.08789 &     4.158 \\ 
Ganymede &   2007 &    4 &   13 &    9 &   42 &    53.81 &    300 &     98.21 &    -1.367 &    -3.698 &    0.08789 &     4.158 \\ 
Vesta &   2007 &    9 &   30 &    0 &   42 &     4.33 &    600 &    101.90 &    -2.703 &     6.091 &    0.08688 &     4.085 \\ 
\object{Pallas} &   2007 &    9 &   30 &    2 &   48 &    24.68 &   3600 &     96.39 &    -1.171 &     1.856 &    0.08666 &     4.005 \\ 
Ceres &   2007 &   12 &   31 &    2 &   57 &     2.45 &   1200 &     96.04 &    -1.290 &     3.243 &    0.08799 &     4.556 \\ 
\object{Iris} &   2008 &    3 &   16 &    7 &   27 &    12.05 &   1200 &     95.41 &    -0.872 &    -1.554 &    0.08809 &     3.619 \\ 
Ceres &   2009 &    1 &   10 &    8 &   10 &    56.32 &    600 &     89.58 &    -1.224 &    -2.815 &    0.08851 &     3.740 \\ 
Ceres &   2009 &    2 &    9 &    7 &    4 &    39.81 &    600 &     94.31 &    -0.885 &    -1.232 &    0.08822 &     4.505 \\ 
Ceres &   2010 &    4 &   29 &    9 &    0 &    52.16 &    600 &    107.85 &    -1.170 &    -2.987 &    0.08766 &     4.816 \\ 
Ceres &   2010 &    4 &   29 &    9 &   15 &    18.70 &    600 &    106.23 &    -1.170 &    -2.988 &    0.08766 &     4.816 \\ 
Vesta &   2011 &   10 &    1 &    0 &   48 &    18.96 &    600 &     98.10 &    -1.482 &     2.994 &    0.09273 &     7.851 \\ 
Vesta &   2011 &   10 &    1 &    3 &   53 &    38.74 &    600 &    102.73 &    -1.483 &     3.004 &    0.09273 &     7.851 \\ 
Vesta &   2011 &   10 &    3 &    1 &   58 &    26.57 &    600 &    102.70 &    -1.526 &     3.105 &    0.09455 &     8.225 \\ 
Vesta &   2011 &   10 &    3 &    4 &    1 &    56.92 &    600 &    104.96 &    -1.543 &     3.111 &    0.09455 &     8.225 \\ 
Vesta &   2011 &   10 &    4 &    1 &   34 &    25.82 &   1200 &     98.90 &    -1.541 &     3.157 &    0.09370 &     8.204 \\ 
Vesta &   2011 &   10 &    4 &    4 &   10 &    20.01 &   1200 &    102.49 &    -1.562 &     3.166 &    0.09370 &     8.204 \\ 
Vesta &   2011 &   10 &    5 &    1 &   31 &    42.97 &    600 &    109.55 &    -1.559 &     3.213 &    0.09361 &     8.194 \\ 
Vesta &   2011 &   10 &    5 &    3 &   57 &    36.27 &    600 &    104.77 &    -1.560 &     3.220 &    0.09361 &     8.194 \\ 
Vesta &   2011 &   11 &    8 &    0 &    5 &    51.54 &    600 &    101.62 &    -2.266 &     5.082 &    0.09776 &    10.728 \\ 
Vesta &   2011 &   11 &    8 &    0 &   26 &    50.22 &    600 &    102.34 &    -2.266 &     5.083 &    0.09776 &    10.728 \\ 
Vesta &   2011 &   12 &    4 &    0 &   18 &    26.23 &    600 &     97.37 &    -2.886 &     6.556 &    0.09562 &     9.073 \\ 
Vesta &   2011 &   12 &    4 &    0 &   28 &    34.80 &    600 &     98.93 &    -2.408 &     6.557 &    0.09562 &     9.073 \\ 
Vesta &   2011 &   12 &    4 &    0 &   39 &    36.85 &    600 &     99.62 &    -2.892 &     6.558 &    0.09562 &     9.073 \\ 
Vesta &   2011 &   12 &    5 &    0 &   18 &    57.23 &    600 &    102.27 &    -2.916 &     6.614 &    0.09558 &     8.397 \\ 
Vesta &   2011 &   12 &    5 &    0 &   29 &    29.80 &    600 &    100.39 &    -2.914 &     6.614 &    0.09558 &     8.397 \\ 
Vesta &   2011 &   12 &    5 &    0 &   40 &     1.79 &    600 &    102.81 &    -2.916 &     6.615 &    0.09558 &     8.397 \\ 
Vesta &   2011 &   12 &    6 &    0 &   29 &    57.34 &    600 &    100.23 &    -2.937 &     6.672 &    0.09553 &     8.429 \\ 
Vesta &   2011 &   12 &    6 &    0 &   40 &    35.35 &    600 &    100.55 &    -2.939 &     6.672 &    0.09553 &     8.429 \\ 
Vesta &   2011 &   12 &    7 &    0 &   19 &    19.62 &    600 &    101.57 &    -2.958 &     6.728 &    0.09549 &     8.512 \\ 
Vesta &   2011 &   12 &    7 &    0 &   30 &     3.83 &    600 &    101.25 &    -2.961 &     6.729 &    0.09549 &     8.512 \\ 
\object{Moon}(245) &   2012 &    6 &    5 &   18 &   42 &    30.31 &  29444$^{*}$ &     99.16 &    -1.541 &    -0.003 &    0.09592 &    10.209 \\ 
Iris &   2012 &    6 &    7 &    1 &    4 &    10.83 &   1800 &    102.92 &    -0.884 &     2.072 &    0.09325 &     8.488 \\ 
Iris &   2012 &    6 &    8 &    3 &   52 &    51.34 &   1800 &     97.60 &    -0.891 &     2.140 &    0.09334 &     8.058 \\ 
\object{Juno} &   2012 &    6 &    8 &    4 &   26 &    19.26 &   1800 &     96.22 &    -0.788 &     1.546 &    0.09325 &     8.488 \\ 
\object{Massalia} &   2012 &    6 &    8 &    8 &   45 &     5.11 &   1800 &     96.13 &    -0.851 &    -1.663 &    0.09592 &    10.209 \\ 
Moon(196) &   2012 &    7 &    5 &   18 &   19 &    51.83 &  18678$^{*}$ &     96.77 &    -1.528 &    -0.005 &    0.09128 &     9.067 \\ 
\object{Melpomene} &   2012 &    7 &    6 &    2 &    0 &    24.95 &   1800 &     98.97 &    -0.836 &     0.877 &    0.09110 &     9.017 \\ 
\object{Io} &   2013 &    2 &   28 &    0 &   36 &    36.21 &    180 &     92.91 &    -6.639 &     6.276 &    0.09126 &     8.506 \\ 
Io &   2013 &    2 &   28 &    0 &   40 &     9.50 &    180 &     93.01 &    -6.654 &     6.276 &    0.09126 &     8.506 \\ 
Io &   2013 &    3 &    1 &    0 &   19 &    40.64 &    180 &     88.00 &    -1.787 &     6.342 &    0.09306 &     8.932 \\ 
Io &   2013 &    3 &    1 &    0 &   23 &    13.97 &    180 &     89.11 &    -1.789 &     6.342 &    0.09306 &     8.932 \\ 
Moon(56) &   2013 &    4 &   28 &   19 &   20 &    51.41 &   6130$^{*}$ &    101.55 &    -1.542 &    -0.007 &    0.09293 &     9.573 \\ 
\object{Irene} &   2013 &    4 &   29 &    3 &   16 &    10.16 &   3600 &     98.42 &    -1.154 &     2.113 &    0.09543 &    11.188 \\ 
Moon(121) &   2013 &    7 &   22 &   20 &   25 &    29.68 &   8030$^{*}$ &     98.61 &    -1.543 &    -0.001 &    0.09394 &     8.452 \\ 
Iris &   2013 &    7 &   23 &    4 &   30 &    39.54 &   1800 &    100.20 &    -1.291 &    -1.298 &    0.09394 &     8.452 \\ 
Massalia &   2013 &   11 &   13 &    3 &   23 &     6.58 &   1800 &    103.17 &    -0.909 &     0.658 &    0.09416 &     9.908 \\ 
& & & & & & & & & & & &\\
  \hline
 \end{tabular}
 ~\\
 $^{*}$ Moon sequence: the number of individual spectra is in brackets and the reported exposure time is the total duration of the sequence. The exposure times of the single measurements range from 30 to 120~s, with an average value of 60~s.
 \label{table1}
 \end{table*}
 \begin{table*}
 \begin{tabular}{lrrrrrrrrrrrr}
 \hline
  & & & & & & & &  & \\
  Object & year & month & day & hour & min & sec & $\tau\,\,$ & RV & $\Delta V_{\rm rel}$ & $t^{\prime}-t$ & Ca~II~K & $<|B|>$  \\
   &  & & &  & & & (s) & (m/s) & (m/s) & (day) & & (Gauss) \\
  & & & & & & & & & & & & \\  
 \hline
 & & & & & & & & & & & & \\
Iris &   2013 &   12 &   19 &    1 &    2 &     2.43 &   1800 &     91.73 &    -2.800 &     6.096 &    0.09368 &     9.032 \\ 
Pallas &   2013 &   12 &   19 &    8 &   40 &    53.76 &   1800 &    104.91 &    -1.486 &    -3.899 &    0.09518 &    12.810 \\ 
Pallas &   2013 &   12 &   20 &    7 &   57 &    34.95 &   1800 &     95.41 &    -1.480 &    -3.851 &    0.09443 &    11.923 \\ 
Pallas &   2013 &   12 &   20 &    8 &   27 &    49.92 &   1800 &     95.37 &    -1.476 &    -3.853 &    0.09443 &    11.923 \\ 
Ganymede(50) &   2014 &    1 &    4 &   17 &   2 &    24.00 &      6000$^{*}$ &    105.57 &    -1.927 &    -0.103 &    0.09706 &    10.787 \\ 
Europa(198) &   2014 &    1 &    6 &   14 &   24 &     0.00 &      11900$^{*}$ &    105.46 &    -2.041 &     0.057 &    0.09890 &    12.927\\ 
Ganymede &   2014 &    2 &   13 &    3 &   22 &    22.93 &     30 &    100.47 &    -1.412 &     2.775 &    0.09331 &    10.038 \\ 
Ganymede &   2014 &    2 &   13 &    3 &   27 &    57.34 &     60 &     97.03 &    -1.414 &     2.775 &    0.09331 &    10.038 \\ 
Ganymede &   2014 &    2 &   13 &    3 &   29 &    28.13 &     60 &    100.28 &    -1.414 &     2.775 &    0.09331 &    10.038 \\ 
Ganymede &   2014 &    2 &   13 &    3 &   31 &     2.42 &     60 &     99.14 &    -1.414 &     2.775 &    0.09331 &    10.038 \\ 
Ganymede &   2014 &    2 &   13 &    3 &   32 &    32.64 &     60 &    100.37 &    -1.415 &     2.775 &    0.09331 &    10.038 \\ 
Ganymede &   2014 &    2 &   13 &    3 &   34 &     6.85 &     60 &     99.50 &    -1.415 &     2.775 &    0.09331 &    10.038 \\ 
Ganymede &   2014 &    2 &   13 &    3 &   35 &    46.92 &     60 &     98.33 &    -1.416 &     2.775 &    0.09331 &    10.038 \\ 
Ganymede &   2014 &    2 &   13 &    3 &   37 &    19.94 &     60 &     98.78 &    -1.416 &     2.775 &    0.09331 &    10.038 \\ 
Ganymede &   2014 &    2 &   13 &    3 &   38 &    49.40 &     60 &     99.58 &    -1.416 &     2.775 &    0.09331 &    10.038 \\ 
Ganymede &   2014 &    2 &   13 &    3 &   40 &    24.40 &     60 &     95.78 &    -1.416 &     2.776 &    0.09331 &    10.038 \\ 
Ganymede &   2014 &    2 &   13 &    3 &   41 &    52.40 &     60 &     98.38 &    -1.416 &     2.776 &    0.09331 &    10.038 \\ 
Europa &   2014 &    2 &   13 &    3 &   50 &    39.40 &     60 &    101.13 &    -1.369 &     2.775 &    0.09331 &    10.038 \\ 
Europa &   2014 &    2 &   13 &    3 &   52 &    13.00 &     60 &     98.53 &    -1.371 &     2.775 &    0.09331 &    10.038\\ 
Europa &   2014 &    2 &   13 &    3 &   53 &    45.80 &     60 &    101.03 &    -1.372 &     2.775 &    0.09331 &    10.038 \\ 
Europa &   2014 &    2 &   13 &    3 &   55 &    15.40 &     60 &     99.43 &    -1.372 &     2.775 &    0.09331 &    10.038 \\ 
Europa &   2014 &    2 &   13 &    3 &   58 &    19.60 &     60 &    103.32 &    -1.376 &     2.776 &    0.09331 &    10.038\\ 
Europa &   2014 &    2 &   13 &    4 &    0 &    22.00 &     60 &    101.62 &    -1.377 &     2.776 &    0.09331 &    10.038 \\ 
Europa &   2014 &    2 &   13 &    4 &    4 &    56.60 &     60 &    101.53 &    -1.366 &     2.776 &    0.09331 &    10.038 \\ 
Moon &   2014 &    2 &   13 &    4 &    8 &    25.60 &     30 &    100.69 &    -1.611 &     0.004 &    0.09344 &    10.475\\ 
Moon &   2014 &    2 &   13 &    4 &   11 &    36.86 &     30 &     99.59 &    -1.611 &     0.004 &    0.09344 &    10.475 \\ 
Pallas &   2014 &    2 &   13 &    5 &    4 &    24.78 &    300 &     98.81 &    -1.275 &    -1.258 &    0.09349 &    10.167 \\ 
Io &   2014 &    4 &   13 &   23 &   24 &    14.16 &    600 &     94.78 &    -1.515 &     7.026 &    0.09611 &    11.647 \\ 
Io &   2014 &    4 &   14 &   23 &   23 &    20.73 &    600 &     87.53 &    -2.370 &     7.100 &    0.09519 &    10.528 \\ 
Io &   2014 &    4 &   14 &   23 &   36 &    15.01 &    600 &     85.90 &    -2.303 &     7.100 &    0.09519 &    10.528 \\ 
Pallas &   2014 &    4 &   15 &    0 &    3 &    58.92 &   1800 &    103.71 &    -2.029 &     2.895 &    0.09877 &    13.290 \\ 
Vesta &   2014 &    4 &   16 &    5 &   50 &    10.76 &    600 &    109.74 &    -0.861 &    -0.565 &    0.09517 &    11.553 \\ 
Pallas &   2014 &    5 &   13 &    1 &    3 &    39.52 &    900 &    101.23 &    -2.548 &     4.448 &    0.09403 &    10.015 \\ 
Vesta &   2014 &    5 &   13 &    4 &    8 &    35.58 &    900 &    101.22 &    -0.976 &     1.612 &    0.09356 &    12.448 \\ 
&  & & & & & & & & & & & \\
 \hline
 \end{tabular}
 ~\\
 $^{*}$ Ganymede or Europa sequence: the number of individual spectra is given in brackets and the reported exposure time is the total duration of the sequence. The exposure times of the single measurements is 60 or 120~s.
 \end{table*}
  \section{Methods}
  \label{methods}
 First we look for periodicities in our solar RV time series by computing its Lomb-Scargle periodogram \citep{HorneBaliunas86}. Then we perform  
 $10\,000$ random permutations of the RV values and measure the height of the highest peak in each of their  periodograms. The frequency of occurrence of a peak equal to or higher than the maximum in the actual time series provides an a posteriori measurement of its false-alarm probability (hereafter FAP), i.e., the probability of its occurrence as a result of the time sampling and the uncorrelated noise present in the dataset. {Another measure of the FAP can be obtained by adding to the individual RV measurements a normally distributed random variable with zero mean and standard deviation equal to the standard deviation of the measurements (cf. Sect.~\ref{rv_uncertainty}), and computing the corresponding periodogram. By repeating this procedure $10\,000$ times, we  estimate the a posteriori  probability of occurrence of a peak of height equal to or greater than the highest peak of the original periodogram.  }
 
 To investigate the correlations between the solar RV and the different activity indexes introduced in Sect.~\ref{activity_index}, we use Spearman  and  Pearson coefficients. The former measures the monotone correlation of one variable vs. the other, while the latter provides a measure of the linear correlation between them. 
{Since the Spearman coefficient is based on the order ranks of the individual variables, it can sometimes have a low absolute value even if they are truly correlated. This happens if one of the variables has a larger scatter for some constant values of the other, that is indeed the case when we consider sequences of RV data acquired on the same night for which the chromospheric and magnetic indexes are constant because they are sampled once per day. In this case, the Pearson coefficient is more suitable to detect a correlation, although the estimate of its significance is much more complex than in the case of the Spearman correlation for which a well-defined theoretical  method is available \citep{Pressetal02}. } 

 We use the {\tt IDL} functions {\tt R\_CORRELATE} and {\tt CORRELATE} to compute the correlation coefficients, respectively. The significance of the obtained values is estimated a posteriori by computing the coefficients for $10\,000$ random permutations of the RV values to evaluate the frequency of obtaining an absolute value equal to or greater than the measured one for an uncorrelated dataset. { Moreover, we test the significance by means of a Monte Carlo simulation of 10 000 data sets obtained by adding random deviates to the original variables. Their standard deviations  are derived in  Sect.~\ref{rv_uncertainty} for the RV and in Sect.~\ref{diff_persp} for the Ca II K and the magnetic flux indexes, respectively.}

 \section{Results}
 \label{results}
 
 \subsection{RV of the individual reflecting bodies}
 \label{individual_bodies}
{ 
For a better characterization of the measurements obtained with the different bodies, we  compute the mean $\langle RV \rangle$ and the standard deviation $\sigma_{\rm RV}$ for the objects having at least 4 measurements. In Table~\ref{individual_objs}, we list the name of the object, the mean radial velocity, the standard deviation, and the number $N$ of measurements. 
\begin{table}
\caption{Solar RV measurements obtained with individual reflecting bodies.}
\begin{center}
\begin{tabular}{lccc}
\hline
Object & $\langle RV \rangle$ & $\sigma_{\rm RV}$ & $ N$ \\
& (m/s) & (m/s) & \\
\hline
& & & \\
Io & 90.18 & 3.36 & 7 \\
Ceres & 98.21 & 6.58 & 7 \\
Vesta & 101.68 & 3.39 & 24 \\
Europa & 101.32 & 3.06 & 10 \\
Ganymede & 99.07 & 1.56 & 17 \\
Pallas & 99.41 & 3.96 & 7 \\
Iris & 97.57 & 4.31 & 5 \\
Moon & 99.39 & 1.67 & 6 \\
&  & & \\
\hline
\label{individual_objs}
\end{tabular}
\end{center}
\end{table}
The mean of all the 88 measurements is 99.196 m/s, their standard deviation is 4.44 m/s and their standard error is 0.47 m/s. Therefore, the mean value of Io is lower by 6.65 standard errors from the mean of all the measurements. This discrepancy is not associated with a specific range of orbital phases of the satellite. However, we find that  Io's measurements were obtained with an airmass ranging from 1.73 to 1.91 with five measurements out of seven taken at an airmass between 1.79 and 1.91. The high airmass and the  variation of the albedo of Io from 0.10 to 0.75 between 380 and 690 nm \citep{NelsonHapcke78} may combine themselves to systematically modify the RV measurements because they reduce the weights of the echelle orders at shorter wavelengths in comparison to those at longer wavelengths which corresponds to a change of the effective line mask used to cross-correlate the spectrum. 

Ceres has the greatest standard deviation, likely as a consequence of its fast rotation ($V_{\rm eq} = 92.3$ m/s) in combination with its surface inhomogeneities (cf. Sect.~\ref{asteroid_rotation}), while Iris has the second largest standard deviation. It depends on one discrepant measurement taken on 2013 December 19 with an airmass of 1.7. After removing that measurement, the mean and the  standard deviation of the remaining 4 measurements of Iris are 99.03 and 3.25 m/s, respectively, in line with the other measurements.  

In conclusion, we decided to discard the measurements obtained with Io and Ceres as well as the discrepant point of Iris and restrict our analysis to the remaining 73 data points. They have a mean RV of  100.26 m/s and a standard deviation of 3.056 m/s. The discarded datapoints  are marked with red dots in Fig.~\ref{timeseries}.  We include in our analysis the measurements obtained with Vesta because they do not show any systematic effect that can affect our results, although its rotational modulation induces a significant scatter as measured by the standard deviation reported in Table~\ref{individual_objs} (cf. Sect.~\ref{asteroid_rotation}). 

The smallest standard deviations are those of the measurements of the Moon -- mostly mean values of sequences of several tens or hundreds of datapoints -- and  of Ganymede that were obtained over two nights. The rotation effects are small in the case of Ganymede and have been corrected in the case of the Moon, thus suggesting that the main source of the standard deviation at the level of $1.6-1.7$ m/s is the intrinsic solar variability. This value is compatible with the 1.33 m/s rms measured by \citet{Dumusqueetal15} during one week between April and May 2014 that fell during a period of moderately high chromospheric activity (cf. Fig.~\ref{solar_cycle}). On a timescale of about two months, \citet{Haywoodetal15} find a mean standard deviation of $\sim 3.9$ m/s  as a consequence of some large activity complexes that modulated the RV with the rotation of the Sun. These measurements encompassed the highest level of chromospheric activity during our time span (see Fig.~\ref{solar_cycle}). The simulations based on the models by \citet{Meunieretal10a} and \citet{Borgnietetal15} give rms of 1.53 m/s during a period of high activity and of 2.62 m/s over the entire cycle. These values are supported by the observations of \citet{Meunieretal10b}. 

We conclude that the difference  between the mean variance of our sequence of 73 points and those of the Moon and Ganymede  is likely due in part to the intrinsic solar activity and in part to the effect of the body rotation through the phenomena introduced in Sect.~\ref{asteroid_rotation}. Given that our observations sample most of activity cycle 24, the difference in the standard deviation of 2.57 m/s appears to be comparable with the intrinsic solar rms expected on that timescale according to the models and the available observations.  }

\subsection{Uncertainty of our RV measurements}
\label{rv_uncertainty}

Given the role of the intrinsic variability of the Sun in determining the standard deviation of our measurements, we expect that it depends on the level of activity of our star. In principle, we can compute the standard deviations of subsets of measurements belonging to intervals of low, intermediate, or high activity to empirically find that dependence. Nevertheless, the presence of a double maximum in cycle 24 and the sparseness of our measurements call for some caution in implementing this approach. 

We subdivide the dataset according to the Ca~II~K index of each point. To have subsets with the same number of points -- that is better to have comparable levels of statistical fluctuation of the standard deviations, given the relatively low numbers of datapoints -- we can consider the measurements with Ca~II~K below and above its median level, that is 0.09334. The first subset consists of 37 datapoints with a mean value of 99.16 m/s and a standard deviation of 2.315 m/s, while the second consists of 36 points with a mean of 101.38 and a standard deviation of 3.338 m/s. The difference between the mean values is significant at 3.3 standard errors indicating that the solar apparent RV increases with its activity level. 

We consider also the case of three intervals with 24, 24 and 25 datapoints in order of increasing mean value of the Ca II K index and find RV mean values of 98.87, 100.51, and 101.35 m/s with standard deviations of 2.460, 2.727, and 3.445 m/s, respectively. The difference between the mean value of the second subset and that of the first is significant at the level of 2.20 standard errors, while the difference between the third and the second is significant only at 0.94 standard errors. 

Considering the case of four subsets with similar numbers of datapoints and increasing mean values of the Ca II K index, we find that the second subset consists of data taken during only one night and therefore it has a standard deviation significantly smaller than the other three subsets.

In conclusion, we  adopt  three subsets to estimate the standard deviations of our measurements as a function of the activity level and assign to each RV measurement the standard deviation of the subset to which it belongs. The effect of the body rotation is made random by the presence of different bodies in each subset, observed at widely different rotation phases, so we do not attempt any correction for it.

\subsection{Periodogram analysis}
{The periodogram of our RV time series is plotted in the top panel of Fig.~\ref{periodogram}. It has the highest peak at 14.196 days with a FAP of 1.18 percent as estimated by $10\,000$ random shuffling of the  RV data. However, if we estimate the FAP by adding random variates with zero mean  and standard deviations equal to the estimated standard deviations of the RV measurements, it increases to 10.08 percent. 

The periodogram of the chromospheric index Ca II K taken with the same cadence is similar to that of the RV variations (Fig.~\ref{periodogram}, middle panel). The peaks at the rotation period and its first harmonic are present also in the periodogram of the complete Ca II K time series where we see concentrations of power close to those periods. Another prominent peak at $\sim 65$~days is present in the chromospheric emission, probably related to large faculae in complexes of activity lasting for 1-2 solar rotations, but it is not particularly conspicuous in the RV periodogram. On longer timescale, the sparseness of our time series strongly affect the periodogram, making it impossible to derive sound results. Nevertheless, we note the broad similarity between the distributions of the relative power of the RV and Ca~II~K index in the  period range $\approx 1-3$ years. 

Although the periodicity at half the solar rotation period has been reported by other  authors \citep[e.g.,][]{Jimenezetal86}, a more continuous sampling is certainly required to confirm its presence in the Sun-as-star RV  time series. 
}
   \begin{figure}
   \centerline{
   \includegraphics[width=10cm,height=11cm,angle=0]{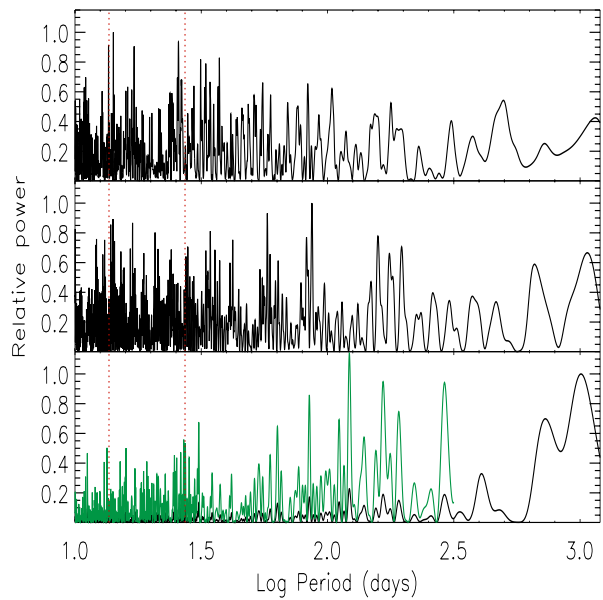}} 
   \caption{Top panel:  the periodogram of our RV time series. Middle panel: the periodogram of the Ca II K index  interpolated on the same calendar as our RV time series. Lower panel: the periodogram of the Ca II K time series with an almost even cadence of one measurement per day. To allow the identification of peaks at short periods, we overplot the periodogram with an amplitude amplified by a factor of five  up to a period of about 300 days (green solid line).   The dotted vertical lines in the three plots mark the solar synodic rotation period and its first harmonic, respectively.  }
              \label{periodogram}%
    \end{figure}
%
 

\subsection{Correlations between the RV and the activity indexes}
 
 We plot in Fig.~\ref{Fig2} our RV measurements vs. the chromospheric Ca II K line index. { The Spearman correlation coefficient for our 73 points is 0.357 with a probability of obtaining an absolute value equal to or greater than it of 0.19 percent in the case of an uncorrelated  dataset, i.e., a significance of 99.8 percent. This result is found by computing the correlation coefficients in the case of $10\,000$ shuffles of the RV time series. On the other hand,  considering $10\,000$ datasets obtained from the observed one by adding normally distributed random variates to the RV and the Ca II K index with their standard deviations, respectively, we find a significance of 95.4 percent. 
 
 The Pearson correlation coefficient is  0.385 with a significance  of 97.2 percent, when it is estimated with $10\,000$ uncorrelated datasets, and of 97.6 percent when it is estimated with  $10\,000$ mock datasets obtained by adding random variates to the RV and chromospheric index with their standard deviations, respectively. }
 
 Since the Spearman coefficient is close to the Pearson coefficient,  a linear model is not too different from the most general monotone correlation model as considered in the computation of the Spearman coefficient.  Therefore, we computed a linear best fit to the RV - Ca II K correlation to estimate the amplitude of the RV variation over the time interval covered by our data and compare it to the variation expected on the basis of the model by \citet{MeunierLagrange13} who showed a linear best fit to their synthetic data in their Fig.~12.  Considering the uncertainties on the RV values, we obtain an RV variation of $4.98 \pm 1.44$~m/s  with the Ca~II~K index ranging between 0.0867 and 0.0990 in the 2006-2014 interval.  This is in good agreement with a variation of $\sim 6.0 \pm 2.5$~m/s as estimated from  Meunier \& Lagrange's work. 
 
 The residuals of the linear best fit to our data range between -5.31 and 9.20~m/s (cf. Fig.~\ref{residuals}) { and a Smirnov-Kolmogorov test shows that they are compatible with a normal distribution with a probability of 0.67.  We derive a standard deviation of 2.82~m/s that can be associated with the effect of the rotation of the reflecting bodies and the variation of solar activity on timescales shorter than the 11-yr cycle (cf. Sects.~\ref{individual_bodies} and~\ref{rv_uncertainty}). }
 
 We plot in Fig.~\ref{Fig3} the RV measurements vs. the absolute value of the mean total magnetic flux measured by  SOLIS VSM. This is the correlation with the highest Spearman and Pearson  coefficients among those obtained with our  two magnetic indexes (cf. Sect.~\ref{activity_index}). { The Spearman coefficient is 0.131, giving an a posteriori significance of only 74.2 percent. On the other hand,  the Pearson coefficient is 0.269 with a significance of 97.8 percent when we derive it from $10\,000$ random shuffles of the RV values. The significance decreases to  84.0 percent when we derive it from $10\,000$ mock datasets obtained by adding random deviates to the two variables. 
 This is probably associated with the limited sensitivity of the Spearman coefficient because of the large spread in several RV measurements having close  values of the total mean magnetic flux (cf. Sect.~\ref{methods}). Therefore, we rely on the Pearson coefficient and conclude that  a correlation is present also between the solar long-term RV variation and the mean total magnetic flux, although with a lower significance than in the case of the RV-Ca II K correlation. The minimum and maximum of the residuals of the linear best fit of the RV-$\langle |B| \rangle $  correlation are $-5.74$ and $9.68$ m/s. They follow a normal distribution with a probability of 0.93 according to the Kolmogorov-Smirnov test and have a standard deviation of 2.94~m/s.  }
   \begin{figure}
   \centerline{
   \includegraphics[width=10cm,angle=0]{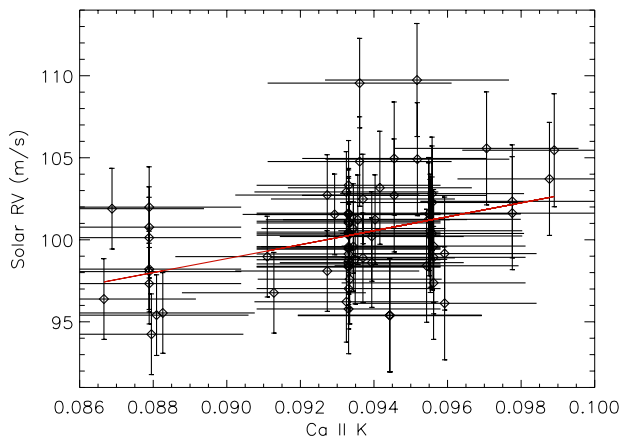}} 
   \caption{Our solar RV measurements vs. the chromospheric Ca II K line index.  The best fit linear regression line is also plotted (red solid line). }
              \label{Fig2}%
    \end{figure}
%
 

   \begin{figure}
   \centerline{
   \includegraphics[width=10cm,angle=0]{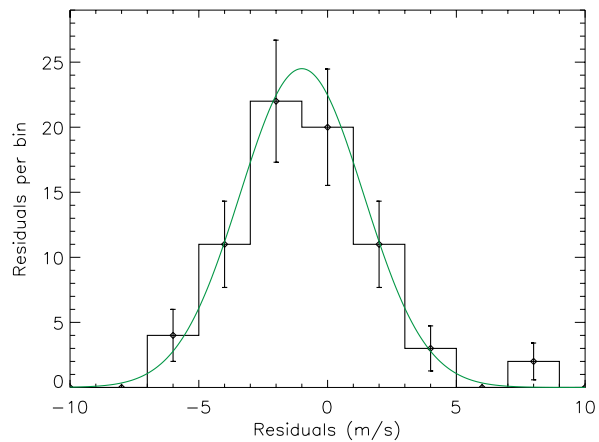}} 
   \caption{Histogram of the distribution of the residuals of the linear best fit to the RV-Ca II K correlation (solid black line). The errorbars are equal to the square root of the number of residuals in each bin. The Gaussian best fit to the distribution is superposed (green solid line).  }
              \label{residuals}%
    \end{figure}
%
 

   \begin{figure}
   \centerline{
   \includegraphics[width=10cm,angle=0]{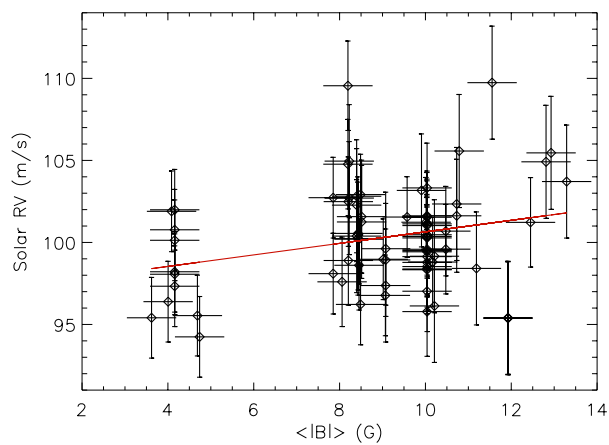}} 
   \caption{Our solar RV measurements vs. the  mean total magnetic flux as measured by SOLIS VSM. The best fit regression line is also plotted (red solid line).  }
              \label{Fig3}%
    \end{figure}
%
 

\section{Discussion and conclusions}

We found a positive correlation between the solar RV variation and the level of chromospheric activity in our star as measured by the Ca II K line index.  A similar correlation is  found between the RV and the absolute value of the mean total photospheric magnetic field of our star, although  with a lower significance. Therefore, the Sun behaves similarly to other low-activity late-type stars that show a positive correlation between the activity level and  the RV variations \citep[e.g., ][]{Lovisetal11,GomesdaSilvaetal12}. 

Our results support the theoretical prediction  that  the quenching of  convective blueshifts associated with localized magnetic fields in the active regions is the main source of the long-term solar RV variations, while the flux perturbations associated with dark spots or bright faculae are of secondary importance \citep{Meunieretal10b,MeunierLagrange13,Dumusqueetal14}. The former effect produces the strongest correlation between RV variation and facular areas because it depends  on their disc-projected area, independently of their position on the solar disc. On the other hand, the flux  effect has a different sign depending on the location of the active regions on the receding or the approaching half of the solar disc. Therefore, a weaker RV-Ca II K correlation and a smaller amplitude of the variation are expected in that case \citep[cf.][]{Lagrangeetal10,Meunieretal10a,Meunieretal10b}. 

The measured amplitude of the long-term RV variation  is in agreement with the models by \citet{MeunierLagrange13}, suggesting that the predictions on the detectability of Earth-size planets  based on their models can be regarded as realistic. Given that the amplitude of  activity cycle 24 was smaller in comparison to those of the  cycles  observed in the 1980s and 1990s, we expect a larger RV variation than  in the 2006-2014 period when the Sun is more active, as would be estimated from, e.g.,  Fig.~12 of \citet{MeunierLagrange13}. 

The results of previous works such as those of \citet{McMillanetal93} or \citet{DemingPlymate94} cannot be directly compared with ours mainly because they used different and limited spectral ranges to measure the solar RV variations. However, the upper limit for the solar RV variation of $\pm\, 4$~m/s in the violet part of the optical spectrum found by \citet{McMillanetal93}  is compatible with our measured range. Contrary to \citet{DemingPlymate94}, we do not detect a highly significant correlation  between the RV variation and the mean magnetic indexes, possibly because they base their measurements on infrared lines that sample a different level of the solar atmosphere where the convective blueshifts may have a different dependence on the magnetic field. 

The correlation with the total magnetic flux density is remarkable during the two solar rotations observed by \citet{Haywoodetal15}  -- they found a Pearson coefficient of 0.58 for 37 nightly-averaged observations --, while it appears to be weaker on our longer timespan. This may be due to the fact that those observations  were taken close to the maximum of cycle 24 at the end of 2011 when the RV variation was dominated by a few relatively large active regions lasting up to two solar rotations. When we consider a longer time span, the intrinsic evolution of many smaller active regions and the size-dependent and non-linear relationship between the convective shift and the intensity of the magnetic field \citep[cf. Fig.~6  in][]{Meunieretal10b} make the correlation with the mean magnetic flux noisier. On the other hand, the correlation with the chromospheric flux appears to be tighter probably because the chromospheric emission is linearly related to the amplitude of the local convective motions that produce the heating of the chromosphere through magnetosonic waves, independently of any size effect that can affect the relationship with the magnetic field. Moreover, \citet{Haywoodetal15} could remove most of the variability induced by the rotation of Vesta, while our correlations are affected by the additional noise due to the rotation of our reflecting bodies. 

{Although a detailed analysis of the solar activity along cycle 24 is beyond the scope of the present work, we notice the difference between the amplitudes of the two relative maxima of the  Ca~II~K and of the magnetic flux  $\langle |B| \rangle$ at the end of 2011 and  in 2014 in Fig.~\ref{timeseries} (cf. middle and lower panels). The variation of the chromospheric index is more similar to the RV modulation (cf. Fig.~\ref{timeseries}, upper panel, black dots), which can explain why it is better correlated with its variation than with the mean magnetic flux. 
Looking at the index timeseries in Fig.~\ref{solar_cycle}, we see the difference in the relative maxima of the mean magnetic flux that reaches a higher level in 2014. Conversely, the monthly sunspot number shows a more prominent and narrow peak in correspondence with the maximum at the end of 2011 \citep[cf. Fig~1 in][]{Kiessetal14}, thus suggesting a redistribution of magnetic flux from larger to smaller structures in the solar photosphere between the two maxima. Since the relationship between the chromospheric emission and the photospheric field depends on the size and type of structures \citep[e.g.][and references therein]{Loukitchevaetal09}, that may account for the difference between the time series of the two activity indexes.  }

The residuals of the best fit to the RV variations in Fig.~\ref{Fig2} range between $-5.31$ and $9.20$ m/s and cannot be attributed only to the effect of the rotation of the asteroids or the Jupiter satellites. The latter may account for an amplitude up to $5-6$ m/s, while the remaining variations are certainly intrinsic to the Sun. Variations with such an amplitude cannot be produced by the line profile distortions induced by surface brightness inhomogeneities, either sunspots or faculae, because their peak amplitude is $\pm \, 2$~m/s during the maximum of the activity cycle \citep[cf.][]{Lagrangeetal10}. The  reduction of convective blueshifts in the active regions can produce an amplitude larger by a factor of $3-4$ that may account for the observed scattering on timescale shorter than $\sim 100$ days that we see in Fig.~\ref{timeseries} \citep[][]{Meunieretal10a}. However, our time sampling is too sparse to reach definite conclusions on the cause of the short-term variations. The work by \citet{Haywoodetal15} addresses this point by means of a daily sampling extended for a couple of solar rotations and more information is expected to come by forthcoming studies \citep[cf.][]{Dumusqueetal15}. 

\begin{acknowledgements}
The authors are grateful to an anonymous referee for several valuable comments and suggestions that helped them to improve their work. They wish to thank Prof.~Andrew~Collier~Cameron and Dr.~Christophe Lovis for interesting discussions. AFL acknowledges support from the National Institute for Astrophysics (INAF) through the {\it Progetti premiali} funding scheme of the Italian Ministry of Education, University and Research. RDH was supported by STFC studentship grant ST/J500744/1 during the course of this work. RDH gratefully acknowledges a grant from the John Templeton Foundation. The opinions expressed in this publication are those of the authors and do not necessarily reflect the views of the John Templeton Foundation.

This work utilizes Ca~II~K solar data obtained at Sacramento Peak Observatory and SOLIS data obtained by the NSO Integrated Synoptic Program (NISP), managed by the National Solar Observatory, which is operated by the Association of Universities for Research in Astronomy (AURA), Inc. under a cooperative agreement with the National Science Foundation.
\end{acknowledgements}

%
%

\appendix
\section{Correction for the Doppler shifts}
\label{appAA}

The HARPS DRS returns the radial velocity $RV_{\rm p}$ relative to the Solar System barycentre by taking into account the apparent position of the target. Therefore, we subtract the barycentric radial velocity correction, called $BERV$, which is recorded in the fits headers, to compute the RV relative to the observer on the Earth. This includes the Doppler shifts associated with the motions of the observer and the reflecting target.  
The Doppler shift $z$ is measured by  the relative line shift: $z \equiv (\lambda_{\rm obs} -\lambda_{0})/\lambda_{0}$, where $\lambda_{\rm obs}$ is the observed wavelength and $\lambda_{0}$ the rest-frame wavelength. The relativistic formula for the Doppler shift is:
\begin{equation}
1 + z = \frac{1 + v_{\rm r}/c}{\sqrt{1 - v^{2}/c^{2}}},
\label{reldoppler}
\end{equation} 
where $v_{\rm r}$ is the radial velocity of the source with respect to the observer, $v$ its total velocity, and $c$ the speed of light \citep[e.g.,][ Eq.~11]{LindegrenDravins03}. In our case, Eq.~(\ref{reldoppler}) will be applied first to the light received by the reflecting target from the Sun and then to the light received by the observer on Earth from the target. Indicating the wavelength received at the target with $\lambda_{\rm a}$ and the rest wavelength at the Sun with $\lambda_{\rm 0}$, we have:
\begin{equation}
\frac{\lambda_{\rm a}}{\lambda_{0}} = \frac{1 + v_{\rm as}^{(r)}/c}{\sqrt{1-v_{\rm as}^{2}/c^{2}}},
\label{ae1}
\end{equation}
where $v_{\rm as}$ is the  velocity of the target with respect to the Sun and $v_{\rm as}^{(r)}$ its radial component along the target-Sun direction, as given by {\tt VmagSn} and {\tt rdot} of NASA Horizon Ephemerides, respectively. The wavelength  $\lambda_{\rm obs}$ measured by a geocentric observer, is given by:
\begin{equation}
\frac{\lambda_{\rm obs}}{\lambda_{\rm a}} = \frac{1 + v_{\rm oa}^{(r)}/c}{\sqrt{1-v_{\rm oa}^{2}/c^{2}}},
\label{ae2}
\end{equation}
where $v_{\rm oa}$ is the  velocity of the target with respect to the centre of the Earth and $v_{\rm oa}^{(r)}$ its radial component,  as given by {\tt VmagOb} and {\tt deldot} of NASA Horizon Ephemerides, respectively. 

The total relativistic change in wavelength $\lambda_{\rm obs}/\lambda_{\rm 0}$ is computed by multiplying Eq.~(\ref{ae2}) by~(\ref{ae1}). 
Therefore, the RV relative to the barycentre of the Sun is obtained as: $RV = RV_{\rm p} - BERV - c z_{\rm rel}$, 
where $z_{\rm rel} \equiv (\lambda_{\rm obs}/\lambda_{0}) -1$.  In other words, the Special Relativity correction to the classic Doppler shift is:
\begin{equation}
\Delta V_{\rm rel} =  \left(v_{\rm as}^{(r)} + v_{\rm oa}^{(r)} \right) - c z_{\rm rel}. 
\label{ae3}
\end{equation}
For the sake of clarity, $\Delta V_{\rm rel}$  is reported in the right column of Table~\ref{table1}.  The correction due to the rotation of the Earth is added classically because its relativistic expression differs from the classic one by less than 1~cm/s in our case.

\end{document}